**Evidence of surface catalytic effect on cosmic dust grain analogues: the ammonia and carbon dioxide surface reaction**

Alexey Potapov[1], Patrice Theulé[2], Cornelia Jäger[1], and Thomas Henning[3]

[1]*Laboratory Astrophysics Group of the Max Planck Institute for Astronomy at the Friedrich Schiller University Jena, Institute of Solid State Physics, Helmholtzweg 3, 07743 Jena, Germany, email: alexey.potapov@uni-jena.de*
[2]*Aix-Marseille Université, CNRS, CNES, LAM, Marseille, France*
[3]*Max Planck Institute for Astronomy, Königstuhl 17, D-69117 Heidelberg, Germany*

**Abstract**

Surface chemistry on cosmic dust grains plays an important role in the formation of molecules at low temperatures in the interstellar and circumstellar environments. For the first time, we experimentally put in evidence the catalytic role of dust surfaces using the thermal reaction $CO_2 + 2NH_3 \rightarrow NH_4^+NH_2COO^-$, which is also a proxy of radical-radical reactions. Nanometre-sized amorphous silicate and carbon grains produced in our laboratory were used as grain analogues. Surface catalysis on grains accelerates the kinetics of the reaction studied at a temperature of 80 K by a factor of up to 3 compared to the reaction occurring in the molecular solid. The evidence of the catalytic effect of grain surfaces opens a door for experiments and calculations on the surface formation of interstellar and circumstellar molecules on dust. Ammonium carbamate on the surface of grains or released intact into protostellar or protoplanetary disk phases can give start to a network of prebiotic reactions. Therefore, there should be a great interest to search for ammonium carbamate and its daughter molecule, carbamic acid, in interstellar clouds, protostellar envelopes, and protoplanetary disks.

**Key words:** astrochemistry – dust, extinction – methods: laboratory: solid state – molecular processes – techniques: spectroscopic

# 1. Introduction

Astrochemistry is a hot topic in the last decades due to the detection of complex organic molecules (COMs) in astrophysical environments, such as cold interstellar molecular clouds and hot protostellar molecular cores [see, for example (Kaifu et al. 1974; Bacmann et al. 2012; Vastel et al. 2014; Potapov et al. 2016)], as well as in comets and meteorites (Cronin & Chang 1993; Elsila et al. 2007; Altwegg et al. 2017). One of the hypothesis about the source of organic compounds, which could serve as the basis of life on Earth, is their formation in the interstellar medium (ISM) and delivery to early Earth on board of meteorites (Oro 1961; Cronin & Chang 1993; Brack 1999; Pearce et al. 2017). Laboratory studies (Bernstein et al. 2002; Muñoz Caro et al. 2002; Nuevo et al. 2006; Elsila, et al. 2007; Nuevo, Milam, & Sandford 2012) have supported this idea and showed that spontaneous generation of amino acids and nucleobases in the ISM is possible in molecular ices covering the surface of cosmic dust grains.

Dust grains play a central role in the physics and chemistry of practically all astrophysical environments. They influence the thermal properties of the medium by absorption and emission of stellar light, provide surfaces for very efficient chemical reactions responsible for the synthesis of a major part of important astronomical molecules from simple ones such as $H_2$ and $H_2O$ to COMs, and they are building blocks of planets. Functional groups and atoms at the surface of dust grains can participate directly in surface reactions and the grain surface itself has a threefold catalytic effect: (i) it provides a place where molecules can rapidly diffuse and react, (ii) it provides a third body to dissipate the energy released in exothermic bond formations, (iii) it lowers the activation barriers of reactions. The first two effects have been shown by a vast number of laboratory experiments performed in interstellar ice analogues covering standard laboratory substrates such as gold, KBr or HOPG, which are not characteristic of cosmic dust grains [for reviews see (Theule et al. 2013; Linnartz, Ioppolo, & Fedoseev 2015; Öberg 2016)]. There is only a handful of studies demonstrating the formation of CO, $CO_2$, and $H_2CO$ molecules involving surface atoms of carbonaceous grains [see (Potapov et al. 2017) and references therein]. The same applies to the catalytic effect of dust surfaces on surface reactions in ices or on bare grains. Only a few studies were performed on the formation of molecular hydrogen and water on silicate surfaces [see (Gavilan et al. 2014; He & Vidali 2014) and references therein]. Water ice can also participate in the sequence of elementary reactions leading to the formation of COMs by lowering one or more reaction barriers (Danger et al. 2014; Noble et al. 2014).

To evaluate the catalytic effect of the dust surface we use the model reaction $CO_2 + 2NH_3 \rightarrow NH_4^+NH_2COO^-$ in the sub-monolayer regime. The two reactants, $CO_2$ and $NH_3$, are two of the several main components of interstellar and cometary ices (Tielens 2013). The reaction is purely thermal and leads to the formation of ammonium carbamate, $NH_4^+NH_2COO^-$, and carbamic acid, $NH_2COOH$, at temperatures higher than 70 K (Bossa et al. 2008; Noble, et al. 2014; Ghesquiere et al. 2018).

## 2. Experimental part

The experiments have been performed using the new ultra-high vacuum set-up, INterStellar Ice Dust Experiment (INSIDE), recently developed in Jena and presented in details elsewhere (Potapov, Jäger, & Henning submitted to ApJ). The base pressure in the main chamber of the set-up is a few $10^{-11}$ mbar allowing to reproduce the physical conditions in dense interstellar clouds and to perform clean experiments without additional adsorption of species such as water from the chamber volume.

In brief, nm-sized, amorphous carbon ($^{13}C$) and silicate ($MgFeSiO_4$) grains were produced in a laser ablation set-up (Jäger et al. 2008; Potapov et al. 2018) by pulsed laser ablation of a graphite or Mg:Fe:Si target and subsequent condensation of evaporated species in a quenching atmosphere of a few mbar He and $H_2$ for carbon grains or He and $O_2$ for silicate grains. Condensed grains were deposited onto a KBr substrate. The thickness of the deposit was 150 nm for $MgFeSiO_4$ (density 3.7 g cm$^{-3}$) and 70 nm for carbon (density 1.7 g cm$^{-3}$) controlled by a microbalance (sensitivity 0.1 nm) using known values for the deposit area and density. A reference surface of pure KBr was also used.

After the deposition of grains, the samples were extracted from the deposition chamber (exposed to air), fixed on a sample holder, and inserted into the pre-chamber of INSIDE, where they were annealed at 200º C during two hours to remove possible surface contamination by molecular adsorbates. After annealing, the sample holder was moved to the UHV chamber of INSIDE and fixed on the coldhead. The pure KBr substrate used in this study was also annealed. $NH_3$ and $CO_2$ (purity 5.5) with a ratio of 4:1 were deposited on the surfaces of KBr, $^{13}C$ grains or $MgFeSiO_4$ grains at 15 K. The gases were introduced through two separated gas lines. The $CO_2$ and $NH_3$ ice thicknesses were calculated from their vibrational bands at 2342 and 1073 cm$^{-1}$ using the band strengths of $7.6\times10^{-17}$ cm molecule$^{-1}$ (Gerakines et al. 1995) and $1.7\times10^{-17}$ cm molecule$^{-1}$ (d'Hendecourt & Allamandola 1986) and assuming a monolayer column density of $1\times10^{15}$ molecules cm$^{-2}$ and a monolayer thickness of 0.3 nm.

After the deposition of ice, the samples were heated up to 80 K with a heating rate of 100 K min$^{-1}$, the initial time was set to zero, and the isothermal kinetics experiments were performed. The temperature was chosen as high enough to perform the efficient reaction and low enough to prevent an active desorption of $CO_2$ and $NH_3$. The formation of the $NH_4^+NH_2COO^-$ product was monitored by the growth of its characteristic vibrational bands at 1550 and 1385 $cm^{-1}$ as a function of time. The column density of $NH_4^+NH_2COO^-$ was calculated from the 1550 $cm^{-1}$ band using the band strength of $2\times10^{-17}$ cm molecule$^{-1}$ (Bossa, et al. 2008). After the kinetics experiments were completed, temperature-programmed desorption (TPD) experiments with a heating rate of 2 K min$^{-1}$ were realized in the temperature range of 80 – 300 K. IR spectra were measured using an FTIR spectrometer (Vertex 80v, Bruker) in the transmission mode. Mass spectra were measured by a quadrupole mass spectrometer (HXT300M, Hositrad) equipped with a multiplier detector (multiplication factor of $10^3$). We followed $CO_2$ on m/z 44, $NH_3$ on m/z 17 and ammonium carbamate on m/z 61 and 78.

The morphology of carbon and silicate grains can be understood as a porous layer of rather fractal agglomerates. The porosity of grains, which can be as high as 90% (Sabri et al. 2014), means an existence of a very large surface. In Figure 1, we show a high-resolution electron microscopy image of amorphous silicate grains, where a high porosity and a large surface are clearly observable. The inset shows carbon grains deposited on a KBr substrate with a nominal surface area of 1 $cm^2$. It was recently shown that 130 nm of water ice mixed with carbon grains and deposited on a nominal substrate surface of 1 $cm^2$ corresponding to a theoretical coverage of about 400 monolayers (ML) desorb following the first order kinetics of desorption (Potapov, Jäger, & Henning 2018). This result was explained by the desorption of water molecules from a large surface of clusters composed of carbon grains. Therefore, the real surface of dust grains deposited on a substrate can be two orders of magnitude larger than the nominal substrate surface of 1$cm^2$.

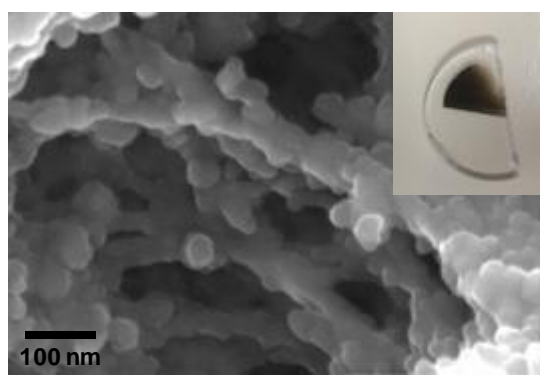

Figure 1. High-resolution electron microscopy image of silicate grains (real surface area is unknown). Inset: a layer of carbon grains on a KBr substrate (nominal surface area is 1 $cm^2$),

To investigate the effect of the surface, it is important to know the monolayer-multilayer transition on the porous carbon and silicate grains we use. The thickness corresponding to the monolayer-multilayer transition is given by the transition between the zeroth-order and first-order form of the Polanyi-Wigner equation (Polanyi & Wigner 1928). According to our TPD curves for ammonia and carbon dioxide taken after the kinetics experiments were completed, we can conclude that the monolayer-multilayer transition takes place on the grains at the ice thickness (calculated from the vibrational bands of $CO_2$ and $NH_3$ as described above) about 200 nm.

### 3. Results

In Figure 2, one can see an example of the IR spectra taken after the deposition of a $CO_2$:$NH_3$ mixture at 15 K and after 4 hours of the isothermal kinetics experiment at 80 K. A detailed assignment of the IR absorption bands of $NH_3$:$CO_2$ ices is done elsewhere (Bossa, et al. 2008; Noble, et al. 2014). Two bands related to $NH_4^+NH_2COO^-$ at 1550 and 1385 $cm^{-1}$ are immediately observable in the IR spectra after heating of the samples up to 80 K.

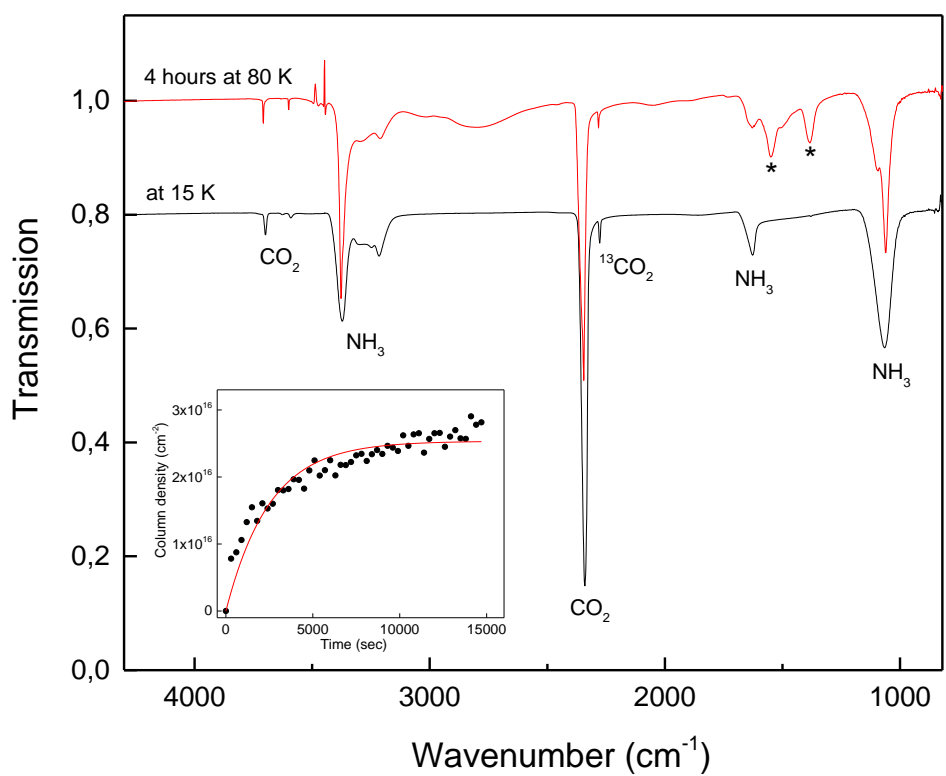

Figure 2. IR spectra taken after the deposition of a $NH_3$:$CO_2$ 4:1 mixture (200 nm thickness) on carbon grains at 15 K and after 4 hours of the title reaction at 80 K. Two bands related to $NH_4^+NH_2COO^-$ are marked by asterisks. The 15 K spectrum is vertically shifted for clarity.

Inset: The time dependence of the $NH_4^+NH_2COO^-$ column density calculated from the 1550 cm$^{-1}$ band derived from the isothermal kinetic experiments performed with $NH_3:CO_2$ 4:1 ices (200 nm thickness) on carbon grains at 80 K.

An example of the results of isothermal kinetics experiments at 80 K is shown in the inset of Figure 2. The exponential decay of $CO_2$ and the corresponding growth of $NH_4^+NH_2COO^-$ exhibit pseudo first-order reaction kinetics (Noble, et al. 2014). For the determination of the rate coefficient we followed the procedure described in (Noble, et al. 2014) solving the kinetic equations:

$$(CO_2)(t) = (CO_2)_0 \times e^{-kt}$$

$$(NH_4^+NH_2COO^-)(t) = (CO_2)_0 \times (1 - e^{-kt})$$

Pseudo-rate coefficients obtained for different surfaces and ice thicknesses are presented in Table 1 and in Figure 3. The reaction rate coefficients presented were obtained from isothermal curves following the formation of ammonium carbamate. Using isothermal curves following the $CO_2$ signal, the reaction rates on grains for ices thinner than 200 nm were found to be higher (~10%) than the rates obtained from the ammonium carbamate curves indicating a slight desorption of $CO_2$ (mainly trapped in the $NH_3$ matrix) in thin ices on grains.

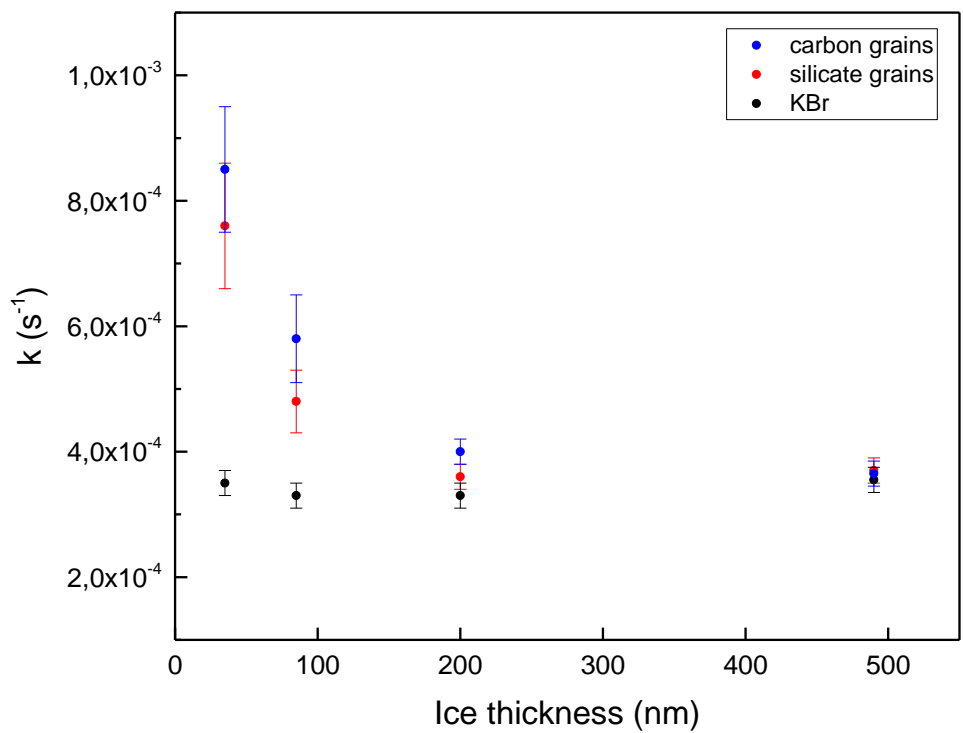

Figure 3. Dependence of the pseudo first-order reaction rate coefficients of the $CO_2 + 2NH_3 \rightarrow NH_4^+NH_2COO^-$ reaction on the ice thickness for various surfaces.

Table 1. Reaction rate coefficients (in $10^{-4}$ s$^{-1}$) for different surfaces and ice thicknesses.

| Ice thickness (nm) | KBr | Carbon grains | Silicate grains |
|---|---|---|---|
| 35 | 3.5 ± 0.2 | 8.5 ± 1.0 | 7.6 ± 1.0 |
| 85 | 3.3 ± 0.2 | 5.8 ± 0.7 | 4.8 ± 0.5 |
| 200 | 3.3 ± 0.2 | 4.0 ± 0.2 | 3.6 ± 0.2 |
| 490 | 3.6 ± 0.2 | 3.6 ± 0.2 | 3.7 ± 0.2 |

After the completion of the kinetic studies, TPD measurements in the temperature range of 80 – 300 K provided signals at masses (m/z) 44 and 17 corresponding to the desorption of $CO_2$ and $NH_3$ at about 95 and 105 K and signals at masses 61 ($NH_2COOH$), 44 ($CO_2$), and 17 ($NH_3$) related to the desorption and fragmentation of $NH_4^+NH_2COO^-$. The signal at mass 78 corresponding to the intact $NH_4^+NH_2COO^-$ molecule was not observed. In Figure 4, we present TPD curves for $NH_4^+NH_2COO^-$ obtained from the mass spectra by taking the signal at mass 61 and from the IR spectra by taking the first temperature derivative of the integrated intensity of the 1550 cm$^{-1}$ band. The curves were measured with a ramp rate of 2 K/min for 200 nm ice on carbon grains after 4 hours of the isothermal kinetic experiment. The results are in agreement with previous experiments (Noble, et al. 2014).

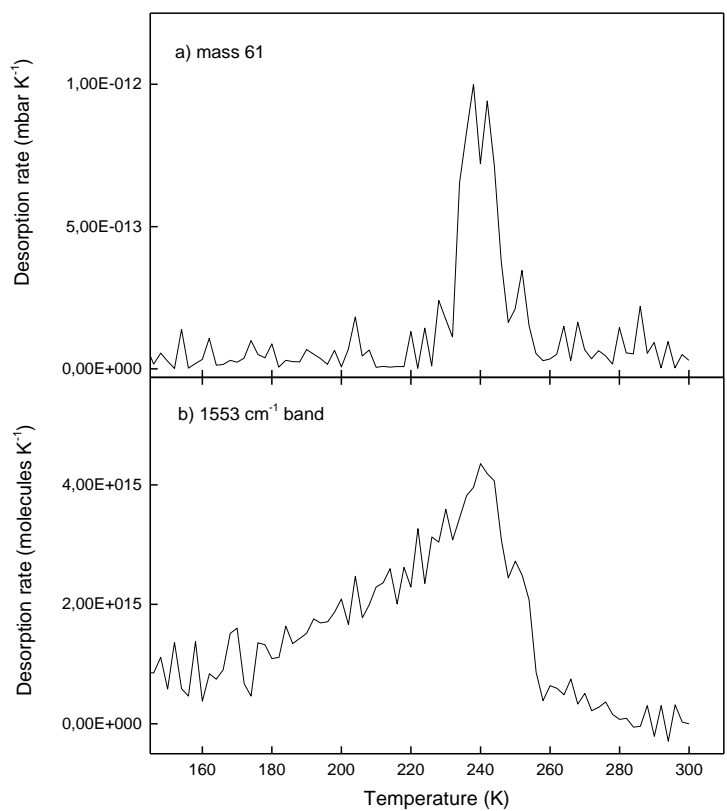

Figure 4. Experimental TPD curves for $NH_4^+NH_2COO^-$ obtained a) from the mass spectra by taking the signal at mass 61 and b) from the IR spectra by taking the first temperature derivative of the integrated intensity of the 1550 cm$^{-1}$ band. The curves were measured with a ramp rate of 2 K/min for 200 nm ice on carbon grains after 4 hours of the isothermal kinetic experiment.

## 4. Discussion

The reaction rate coefficients obtained in the present study are equal for all surfaces when the ice thickness is about 500 nm corresponding to more than 1500 ML coverage. The reaction rate coefficient for the reaction on KBr stays constant in the limits of uncertainties in the whole range of ice thicknesses. However, with the decrease of the ice thickness, starting from 200 nm thickness, we observed a dramatic increase of the reaction rate coefficient reaching values close to $1\times10^{-3}$ s$^{-1}$ at an ice coverage of about 35 nm or 100 ML. Thus, we measured up to three times higher reaction rates on grains compared to the molecular solid on KBr depending on the ice thickness. If we compare with the literature value $4.9\times10^{-5}$ s$^{-1}$ (Noble, et al. 2014) measured at 80 K for a $NH_3$:$CO_2$ 4:1 mixture in a molecular solid, the reaction rate on grains is up to 20 times higher. The difference between our KBr values and the literature value could be explained by different experimental procedures of two studies.

The increased reaction rates for grains compared to those measured for $NH_3$:$CO_2$ molecular solids can be explained in two ways: (i) an increased diffusion rate of species and (ii) a participation of dust surfaces in the reaction studied.

The first explanation is purely physical: at low temperature the reaction-diffusion is diffusion-limited. The diffusion of the reactant molecules on a surface is much faster than in a bulk material (Mispelaer et al. 2013; He, Emtiaz, & Vidali 2018). It has been shown that the diffusion of $CO_2$ and $NH_3$ reactants in the volume of ice occurs along the surface of cracks generated by the structural changes in water ice mantles upon crystallization, which leads to a dramatic increase of the production rate of ammonium carbamate as compared to what it can be with bulk diffusion (Ghesquiere, et al. 2018). In our case of pure $NH_3$:$CO_2$ solid, the surface is provided by aggregates/clusters of nm-sized carbon and silicate nm-sized grains seen in Figure 1.

The second explanation implies a participation of the dust surface in the $CO_2$ + 2$NH_3$ reaction. First, the reaction $CO_2$ + 2$NH_3$ + dust can lead to the formation of an intermediate weakly bound $CO_2$–$NH_3$ complex, which could be stabilized by a third body (dust surface) that helps to overcome the reaction barrier. The formation of weakly bound complexes on the surface of grains in the ISM is a matter of a number of experimental and theoretical studies [see (Potapov 2017) for a brief review]. Secondly, the reaction barrier of the $CO_2$ + 2$NH_3$ reaction can be different for different surface compositions. It can vary by several kJ mol$^{-1}$ depending on the concentration ratio of the $NH_3$:$CO_2$:$H_2O$ cluster (Noble, et al. 2014). To the best of our knowledge, no calculations on carbon or silicate surfaces exist.

## 5. Astrophysical implication

Sub-monolayer reactions are important in two astrophysical stages or environments: (i) during the transition from diffuse to dense clouds and the formation of the first monolayer, which usually takes place during the first $10^5$ years, (ii) during the desorption of the volatile mantles and the subsequent concentration of refractory materials taking place at the snowline of protostellar envelopes and protoplanetary disks. The fact that the surface has an effect on the sub-monolayer or on the bottom most layer can result in an enhanced solid-state reactivity for reactions involving molecules and/or radicals.

The reaction products of $NH_3$ and $CO_2$, ammonium carbamate and carbamic acid can store $CO_2$ and $NH_3$ on the surface of dust grains in interstellar clouds. During the protostellar phase, grains are heated and above 250 K, when the carbamic acid and ammonium carbamate desorb, $CO_2$ and $NH_3$ can be introduced into the gas phase, at much higher temperatures than their normal desorption temperatures.

Once delivered on our planet, both ammonium carbamate and carbamic acid can further evolve under the primitive Earth environment. It was discovered that ammonium carbamate can be converted into urea $CO(NH_2)_2$ and water in a temperature range between 135º and 200º C (Clark, Gaddy, & Rist 1933). Formamide and urea was also produced in interstellar ice analogues by UV irradiation of a HNCO ice (Raunier et al. 2004) or a $CH_3OH:NH_3$ ice mixture (Nuevo et al. 2010). We can assume that UV irradiation of "solid" ammonium carbamate may also lead to the formation of urea. Urea is a molecular corner stone in the search for the origin of life (Miller & Urey 1959). It can be a precursor of pyrimidine required for the synthesis of nucleobases in RNA molecules (Robertson & Miller 1995). Urea was tentatively detected in the ISM, on grains as well as in the gas phase (Raunier, et al. 2004; Remijan et al. 2014).

Thus, ammonium carbamate on the surface of grains or released intact into the protostellar or protoplanetary disk phases can give start to a network of prebiotic reactions. The search for ammonium carbamate and its daughter molecule, carbamic acid, in interstellar clouds, protostellar envelopes, and protoplanetary disks should be of great interest. However, the problem is that these molecules are not stable at room temperature in the gas phase. Therefore, high-resolution gas-phase spectra of the molecules, which allow a radio-astronomical search, are not available. One possibility to obtain their spectroscopic passports is to synthesize the molecules in interstellar ice analogues and detect them above the surface of ice immediately after their release into the gas phase. This can be realised by a combination of solid state chemistry and high-resolution gas phase spectroscopy in one experiment. Such an idea is currently under development in a few groups worldwide.

There are a number of important surface thermal reactions for prebiotic astrochemistry, for example $NH_3 + HCN \rightarrow NH_4^+CN^-$ (Noble et al. 2013) and $CO_2 + 2CH_3NH_2 \rightarrow CH_3NH_3^+CH_3NHCOO^-$ (Bossa et al. 2009). $NH_4^+CN^-$ can react with $CH_2NH$, which is a product of HCN hydrogenation, leading to aminoacetonitrile $NH_2CH_2CN$ that can form glycine according to the Strecker synthesis. $CH_3NH_3^+CH_3NHCOO^-$ acts as a glycine salt precursor in VUV environments. In this context, it is also worth to mention hydrogenation reactions such as the hydrogenation of CO leading to the formation of methanol (Hiraoka et al. 1994) and the hydrogenation of HCN leading to the formation of methylamine (Theule et al. 2011), a direct precursor of glycine. All these reactions can occur on a much shorter timescale than it is thought now.

## 6. Summary

The present study demonstrates that the rate coefficients for the thermal reaction $CO_2 + 2NH_3 \rightarrow NH_4^+NH_2COO^-$ on the surface of nm-sized carbon and silicate grains, which are reliable analogues of interstellar and circumstellar dust grains, are up to 3 times higher compared to the rate coefficients measured in the molecular solid. Thus, the catalytic effect of grains is clearly observed and must be taken into account. This result shows the role of the surface in the kinetics of solid-state reactions. A faster kinetics, on this particular reaction or other ones, especially radical-radical reactions, should have a great impact on astrochemical modelling involving surface chemistry.

**Acknowledgments**

We would like to thank two anonymous referees for questions, suggestions, and corrections that helped to improve the manuscript. This work was partly supported by the Research Unit FOR 2285 "Debris Disks in Planetary Systems" of the Deutsche Forschungsgemeinschaft (grant JA 2107/3-1).